\documentclass[aps,prc,showpacs,floatfix,preprint]{revtex4-1}
\usepackage{graphicx}
\usepackage{epsfig}
\usepackage{color}
\usepackage[normalem]{ulem}  

\begin{document}
\title{Effect of the Coulomb interaction on the liquid-gas phase transition of nuclear matter}
 \author{Rana Nandi}
 \email[]{nandi@fias.uni-frankfurt.de}
 \affiliation{Frankfurt Institute for Advanced Studies, 60438 Frankfurt am Main, Germany}
 
 \author{Stefan Schramm}
 \email[]{schramm@fias.uni-frankfurt.de}
 \affiliation{Frankfurt Institute for Advanced Studies, 60438 Frankfurt am Main, Germany}
 
 \begin{abstract}
  We investigate the role of the Coulomb interaction on the liquid-gas phase transition of
  nuclear matter with three different values of proton fraction ($Y_p=0.5,0.3$ and 0.1), relevant for
  heavy-ion physics as well as various astrophysical scenarios, within the framework of
  quantum molecular dynamics. We perform simulations for a wide range of density and temperature
  with and without the Coulomb interaction and calculate the two-point correlation functions
  of nucleon density fluctuations for all the configurations to determine the phase transition region.
  We also determine the critical end point of the liquid-gas phase transition for all three values of proton fraction considered. 
  We observe that  the Coulomb interaction reduces the transition temperature by $\gtrsim 2$ MeV for nuclear matter 
  with $Y_p=0.5$ and $0.3$ and by $\sim 1$ MeV for nuclear matter  with $Y_p=0.1$. However, the critical
  density is found to be more or less insensitive to the Coulomb interaction. 
 \end{abstract}

 \maketitle
\section{Introduction}
One main focus of heavy-ion collision experiments is to understand the properties of the liquid-gas phase transition in
nuclear matter \cite{Agostino05,Das05}. This phase transition is also important for various astrophysical reasons.
For example, it plays a significant role in the dynamics of supernova explosions \cite{Hempel10,Pais15, Ducoin07} and neutron stars
\cite{Sharma12, Typel14, Raduta10, Ducoin07}. There exist numerous studies on the liquid-gas phase-transition of both symmetric and 
asymmetric nuclear matter using non-relativistic Skyrme interactions \cite{Jaqaman84, Lattimer85, Pethick95, Kolomietz01, Ducoin06} as well as 
relativistic mean-field models \cite{Mueler95, Wang00, Avancini06, Pais15,Hempel13}. Studies of the the liquid-gas mixed phase are mostly done using the 
Gibbs phase equilibrium conditions derived in bulk limit i.e. ignoring the finite-size effects due to the surface and Coulomb interactions 
\cite{Mueler95, Avancini06, Ducoin06, Hempel13, Wang00, Wellenhofer15, Pais15}. Several authors have included finite-size effects but at 
different levels of approximations and obtained considerable effects on the liquid-gas phase transition properties 
\cite{Jaqaman84, Lee01, Sil04, Avancini08, Bao14}. Recently, in Ref. \cite{Bao16} the influence of surface and Coulomb interactions on the
liquid-gas phase transition of stellar matter is studied in a consistent manner by using a compressible liquid-drop model where the surface 
and Coulomb contributions are included while deriving the phase equilibrium conditions. They found that the finite-size effects significantly
reduce the region of liquid-gas mixed phase and the critical temperature ($T_c$) is much lower than that obtained with bulk calculation.

In this article, we investigate the influence of Coulomb interaction on the liquid-gas phase transition of 
nuclear matter with quantum molecular dynamics (QMD) simulation. In particular, we use the QMD model developed by Maruyama {\it et al} 
\cite{Maruyama98} and extensively used to study the various properties of pasta phases that appear at the liquid-gas transition region, 
in recent years \cite{Watanabe03, Watanabe04, Watanabe05, Watanabe09, Nandi16}.

\section{Formalism}\label{sec:formalism}
In the QMD approach the state of a nucleon is represented by a Gaussian wave packet (we set $\hbar=c=1$):
\begin{equation}
 \psi_i({\bf r}) = \frac{1}{(2\pi C_W)^{3/4}} \exp\left[-\frac{({\bf r - R_i})^2}{4C_W}+ i \,{\bf r\cdot P_i}\right],
\end{equation}
where ${\bf R}_i$ and ${\bf P}_i$ denote the centers of the position and momentum of the wave packet $i$, respectively, with the corresponding
width $C_W$. Then the total wave function for the ${\cal N}$-nucleon system is obtained by taking the direct product of single-nucleon wave functions
\begin{equation}
 \Psi(\{{\bf r}\}) =  \prod_i^{\cal N} \psi_i({\bf r}) 
\end{equation}
Here we adopt the QMD Hamiltonian developed by Maruyama {\it et al.} \cite{Maruyama98}, to simulate
the nuclear matter at sub-saturation densities. The Hamiltonian consists of several terms :
\begin{equation}
 {\cal H}=T+V_{\rm Pauli}+V_{\rm Skyrme}+V_{\rm sym}+V_{\rm MD}+V_{\rm Coul},
\end{equation}
where $T$ is the kinetic energy, $V_{\rm Pauli}$ is the  Pauli potential, which phenomenologically incorporates the Pauli exclusion
principle, $V_{\rm Skyrme}$ represents the nucleon-nucleon potential similar to Skyrme-like interactions, $V_{\rm sym}$ is 
the isospin-dependent potential related to the symmetry energy, $V_{\rm MD}$ is the momentum-dependent potential included as
Fock terms of Yukawa-type interactions and finally, $V_{\rm Coul}$ is the Coulomb potential. 
The explicit expressions for all the terms are given as \cite{Maruyama98, Nandi16}:
\begin{widetext}
\begin{eqnarray}
  T  &=& \sum_i \frac{\bf P_{\it i}^{2}}{2 m_{i}}\ ,\label{kin}\\  
  V_{\rm Pauli} &=& 
  \frac{C_{\rm P}}{2}\
  \left( \frac{1}{q_0 p_0}\right)^3
  \sum_{i, j(\neq i)} 
  \exp{ \left [ -\frac{({\bf R}_i-{\bf R}_j)^2}{2q_0^2} 
          -\frac{({\bf P}_i-{\bf P}_j)^2}{2p_0^2} \right ] }\
  \delta_{\tau_i \tau_j} \delta_{\sigma_i \sigma_j}\ ,\label{pauli}\\
  V_{\rm Skyrme} &=&
  {\alpha\over 2\rho_0}\sum_{i, j (\neq i)}
  \rho_{ij}
  +  {\beta\over (1+\tau)\ \rho_0^{\tau}}
  \sum_i \left[ \sum_{j (\neq i)} 
                     \tilde{\rho}_{ij}  \right]^{\tau}\ ,
                   \label{skyrme}\\
   V_{\rm sym} &=&
  {C_{\rm s}\over 2\rho_0} \sum_{i , j(\neq i)} \,
  ( 1 - 2 | \tau_i - \tau_j | ) \ \rho_{ij} \\
  V_{\rm MD}  &=&
         {C_{\rm ex}^{(1)} \over 2\rho_0} \sum_{i , j(\neq i)} 
      {1 \over 1+\left[{{\bf P}_i-{\bf P}_j \over \mu_1}\right]^2} 
      \ \rho_{ij}
     +   {C_{\rm ex}^{(2)} \over 2\rho_0} \sum_{i , j(\neq i)} 
      {1 \over 1+\left[{{\bf P}_i-{\bf P}_j \over  \mu_2}\right]^2} 
      \ \rho_{ij}\ ,\label{md}\\
  V_{\rm Coul} &=&
  {e^2 \over 2}\sum_{i , j(\neq i)}
  \left(\tau_{i}+\frac{1}{2}\right) \, \left(\tau_{j}+\frac{1}{2}\right)
  \int\!\!\!\!\int d^3{\bf r}\,d^3{\bf r}^{\prime} 
  { 1 \over|{\bf r}-{\bf r}^{\prime}|} \,
  \rho_i({\bf r})\rho_j({\bf r}^{\prime})\ ,\label{coulomb}
\end{eqnarray}
\end{widetext}
where $\rho_0=0.165$ fm$^{-3}$ is the normal nuclear matter density, $\sigma_{i}$ and $\tau_{i}$ ($1/2$ for protons and $-1/2$ for neutrons)
are the nucleon spin and isospin, respectively and $\rho_{ij}$ and $\tilde{\rho}_{ij}$   represent the overlap between single-nucleon densities 
and are defined as
\begin{equation}
  \rho_{ij} \equiv \int { d^3{\bf r} \rho_i({\bf r})
  \rho_j({\bf r}) }\ ,\quad \tilde{\rho}_{ij} \equiv \int { d^3{\bf r} \tilde{\rho_i}({\bf r})\tilde{ \rho_j}({\bf  r})}\ ,
  \label{rhoij}
\end{equation}
whereas the single-nucleon densities are given by
\begin{eqnarray}
  \rho_i({\bf r}) & = & \left| \psi_{i}({\bf r}) \right|^{2}
  = \frac{1}{(2\pi C_W)^{3/2}}\ \exp{\left[
                - \frac{({\bf r} - {\bf R}_i)^2}{2C_W} \right]}\ ,\quad \\
  \tilde{\rho_i}({\bf r}) & = &
  \frac{1}{(2\pi \tilde{C}_W)^{3/2}}\ \exp{\left[
                - \frac{({\bf r} - {\bf R}_i)^2}{2\tilde{C}_W} \right]}\ , 
 \end{eqnarray}
with
\begin{equation}
  \tilde{C}_W = \frac{1}{2}(1+\tau)^{1/ \tau}\ C_W .
\end{equation}
The modified width $\tilde{C}_W$ of the Gaussian wave packet is introduced to adjust the effect of density-dependent terms 
\cite{Maruyama98}. The Hamiltonian has 12 parameters shown in Table \ref{tab:parameter}. They are determined to reproduce
the saturation properties of nuclear matter as well as ground state properties of finite nuclei.
\begin{table}[]
\caption{Parameter set for the interaction \cite{Maruyama98}}
{\small \begin{tabular}{cccc}
\hline\hline 
& $C_{\rm P}$ (MeV) &\qquad\qquad 207 &\\
& $p_{0}$ (MeV/$c$) &\qquad\qquad 120 &\\
& $q_{0}$ (fm) &\qquad\qquad 1.644 &\\
& $\alpha$ (MeV) &\qquad\qquad $-92.86$ &\\
& $\beta$ (MeV) &\qquad\qquad 169.28 &\\
& $\tau$ &\qquad\qquad 1.33333 &\\
& $C_{\rm s}$ (MeV) &\qquad\qquad 25.0 &\\
& $C_{\rm ex}^{(1)}$ (MeV) &\qquad\qquad $-258.54$ &\\
& $C_{\rm ex}^{(2)}$ (MeV) &\qquad\qquad 375.6 &\\
& $\mu_1$ (fm$^{-1}$) &\qquad\qquad 2.35 &\\
& $\mu_2$ (fm$^{-1}$) &\qquad\qquad 0.4 &\\
& $C_W$ (fm$^2$) &\qquad\qquad 2.1 &\\
\hline\hline 
\end{tabular}}
\label{tab:parameter}
\end{table}

In order to obtain the equilibrium configuration we adopt the QMD equations of motion with damping terms \cite{Maruyama98}:
\begin{eqnarray}
 {\bf\dot{R}_i } &=& \frac{\partial H}{\partial {\bf P_i}} - \mu_R\frac{\partial H}{\partial {\bf R_i}},\nonumber\\ 
 {\bf\dot{P}_i } &=& -\frac{\partial H}{\partial {\bf R_i}} - \mu_P\frac{\partial H}{\partial {\bf P_i}},\label{eom}
\end{eqnarray}
where the damping coefficients $\mu_R$ and $\mu_P$ are positive definite and relate to the relaxation time scale.

As the QMD Hamiltonian used here contains momentum-dependent interactions ($V_{\rm Pauli}$  and $V_{\rm MD}$), we cannot use the
usual expressions for the instantaneous temperature given as :
\begin{equation}
\frac{3}{2}\, T=\frac{1}{\cal N}\sum_{i=1}^{\cal N} \frac{{\bf P}_i^2}{2m_i},
\label{eq:Tkin}
\end{equation}
where ${\cal N}$ is the number of particles. Instead we use the effective temperature defined as \cite{Chikazumi01}:
\begin{equation}
 \frac{3}{2}\, T_{\rm eff} = \frac{1}{\cal N}\sum_{i=1}^{\cal N}\frac{1}{2}{\bf P}_i\cdot \frac{\partial{\cal H}}{\partial{\bf P}_i},
 \label{eq:Teff}
\end{equation}
which reduces to the usual definition of Eq. (\ref{eq:Tkin}) if the Hamiltonian does not contain momentum-dependent interactions.
Performing Metropolis Monte Carlo simulations  it was shown in  Ref. \cite{Watanabe04} that $T_{\rm eff}$ is consistent with the 
temperature in the Boltzmann statistics.

In order to perform simulations at a specified temperature ($T_{\rm set}$) we adopt the Nos\'{e}-Hoover thermostat \cite{Nose84, Hoover85, Allen}
after suitably modifying it to adapt to the effective temperature \cite{Watanabe04}. The Hamiltonian including the thermostat is given by:
\begin{equation}
  {\cal H}_{\rm Nose} = \sum_{i=1}^{\cal N}\frac{{\bf P}_i^2}{2m_i} + {\cal U}(\{{\bf R}_i\},\{{\bf P}_i)\} + \frac{s^2p_s^2}{2}\,
  + g\frac{{\rm ln}\,s}{\beta}
\end{equation}
where ${\cal U} (\{{\bf R}_i\}),\{{\bf P}_i\})  = {\cal H} - T$ is the potential depending on both positions and momenta, 
$s$ is the extended variable for the thermostat, $p_s$ is the momentum conjugate to $s$, $Q$ is the effective ``mass'' associated 
with $s$ taking a value $\sim10^8\, {\rm MeV\, fm}^2$, $g=3{\cal N}$ needed to generate the canonical ensemble, and $\beta=1/T_{\rm set}$.
The equations of motion for the extended system are written as:
\begin{eqnarray}
 {\bf\dot{R}_i} &=& \frac{{\bf P}_i}{m_i} + \frac{\partial{\cal U}}{\partial {\bf P}_i}\,\\
 {\bf\dot{P}_i} &=& -\frac{\partial {\cal U}}{\partial {\bf R_i}} - \xi{\bf P}_i,\\
 \dot{\xi} &=& \frac{1}{Q}\left[\sum_{i=1}^{\cal N}\left(\frac{{\bf P}_i}{m_i}+{\bf P}_i\cdot\frac{\partial{\cal U}}{\partial{\bf P}_i} \right)
 -\frac{g}{\beta}\right]\,\\
 \dot{s}/s &=& \xi \, 
\end{eqnarray}
where  $\xi(=sp_s/Q$) acts as thermodynamic friction coefficient. When the system is evolved according to the above equations
${\cal H}_{\rm Nose}$ remains conserved and $T_{\rm eff}$ fluctuates around $T_{\rm set}$.

\section{Simulation}\label{sec:simulation}
Adopting the theoretical framework outlined in the previous section we perform QMD simulations of nuclear matter
for a wide range of density ($\rho=0.1-0.775\rho_0$) and temperature ($T=1-9$ MeV) relevant for the study of the liquid-gas phase
transition. We investigate symmetric nuclear matter (proton fraction $Y_p=0.5$) important for heavy-ion collisions
as well as asymmetric nuclear matter with $Y_p=0.3$, typical for supernova environment and $Y_p=0.1$, relevant for 
neutron stars. We take  into account 2048 nucleons for $Y_p=$ 0.5 and 0.3, and 16384 nucleons for $Y_p=0.1$ in a cubic box the size of which
is determined from the number of particles and the chosen density. Periodic boundary conditions are imposed to simulate infinite matter. 
The number of protons (neutrons) with spin-up is taken to be equal to that of protons (neutrons) with spin-down. To calculate
the Coulomb interaction we employ the Ewald method \cite{Allen, Watanabe03}, where electrons are considered to form  a uniform background and
make the system charge neutral. To study the nuclear matter at sub-saturation densities several authors \cite{Horowitz04, Schneider13, Dorso12}
have considered the Coulomb interaction as a Yukawa-type interaction where the choice of screening length ($\lambda$) is not very well defined.
However, in a recent study \cite{Alcain14} it was shown that the the results may depend on $\lambda$, significantly. 
The Ewald method used here does not suffer from this shortcoming.

As an initial configuration we distribute nucleons randomly in phase space. Then with the help of the Nos\'{e}-Hoover 
thermostat we equilibrate the system at $T\sim20$ MeV for about $2000$ fm/c. To achieve the ground state
configuration we then slowly cool down the system in accordance with the damped equations of motion (Eqs. \ref{eom}) until the 
temperature reaches a value below $1$ keV. In order to obtain nuclear matter configuration at a finite temperature $T_{\rm set}$
we cool down the system until $T$ reaches $\sim 5$ MeV. Then the system is relaxed for $5000$ fm/c at the desired temperature
$T_{\rm set}$ with the help of the thermostat and finally, it is further relaxed without the thermostat for another 
$5000$ fm/c. All the measurements are taken at this last stage of simulation.

\section{Results}
\begin{figure}
 \begin{tabular}{cc}
  \includegraphics[width=0.45\textwidth]{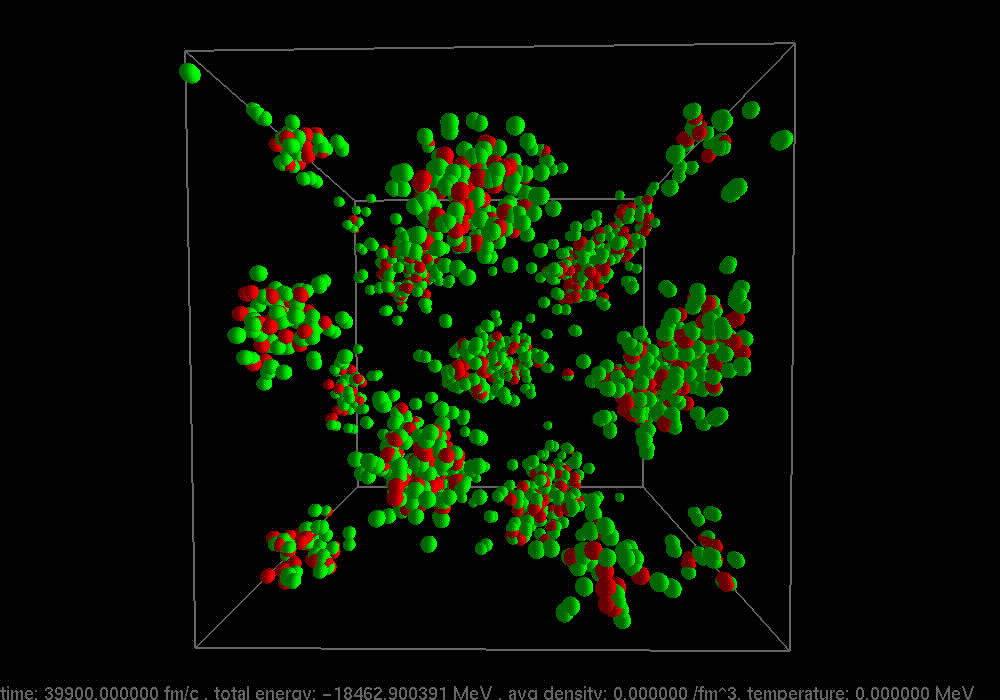}&
  \includegraphics[width=0.45\textwidth]{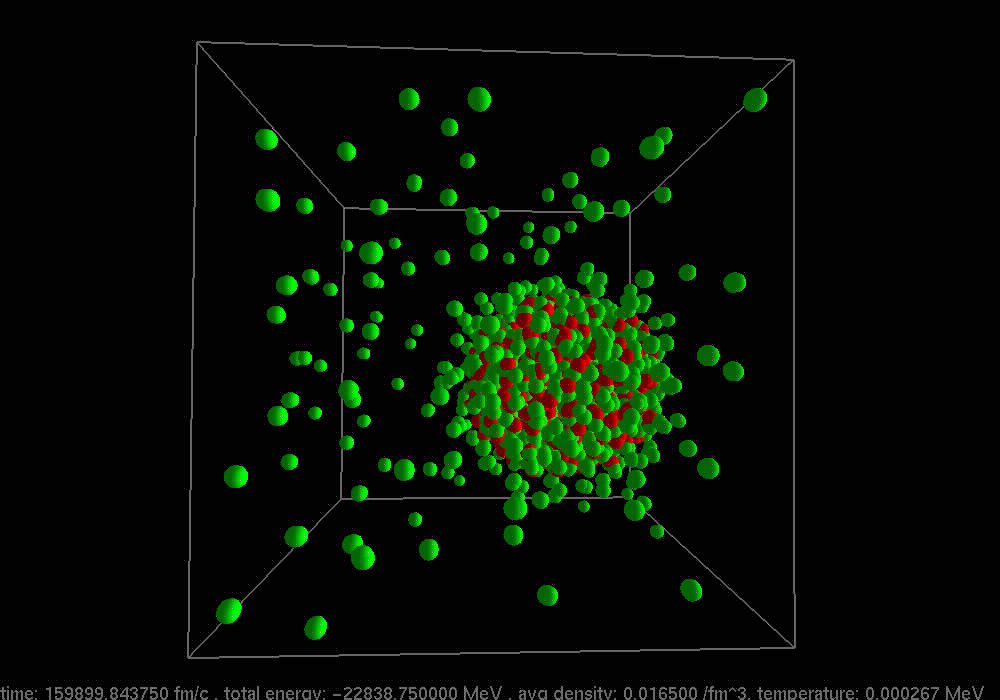}\\
 \end{tabular}
 \caption{Snapshots from simulations showing distribution of nucleons at $0.1\rho_0\,,Y_p=0.3,\, T=0$ with (left) and without (right) 
 the Coulomb interaction. Green (red) spheres represent neutrons (protons). }
 \label{fig:snapshots}
\end{figure}
In Fig. \ref{fig:snapshots} we show simulation snapshots for the nucleon distributions of asymmetric matter with $Y_p=0.3$,
at $\rho=0.1\rho_0$ and $T=0$. The snapshot in the left (right) panel is obtained when the Coulomb interaction is (not) taken into account. 
As expected, we get a single large cluster with several dripped neutrons in absence of the repulsive Coulomb interactions. On the other hand,
in presence of the Coulomb interaction we observe several smaller clusters arranged in a lattice.

For the analysis of nucleon distribution in space we calculate the two-point correlation function $\xi_{ii}$ for the nucleon density 
fluctuation defined as \cite{Watanabe03,Nandi16}:
\begin{equation}
 \bigtriangleup^{(i)} = \frac{\rho^{(i)}({\bf x}) - \rho_{\rm av}^{(i)}}{\rho_{\rm av}^{(i)}}
\end{equation}
where $i=n,p,N$ denotes neutrons, protons and nucleons, respectively and $\rho_{\rm av}^{(i)}={\cal N}^{(i)}/V$.
Then the correlation function is given by
\begin{equation}
 \xi_{ii}(r) = \left<\bigtriangleup_i({\bf x})\bigtriangleup_i({\bf x} + {\bf r})\right>\,,
\end{equation}
where the average is taken over the position $\bf{x}$ and the direction of $\bf{r}$.

\subsection{Symmetric nuclear matter}
To investigate the role of the Coulomb interaction on the liquid-gas phase transition of symmetric nuclear matter we perform simulations
for the whole range of density and temperature mentioned earlier with and without Coulomb interaction. We take 2048 nucleons and obtain
the ground state configurations following the procedure described in the previous section. The phase transition region is determined
by calculating the two-point correlation functions.
\begin{figure}
 \begin{tabular}{cc}
  \includegraphics[width=0.45\textwidth]{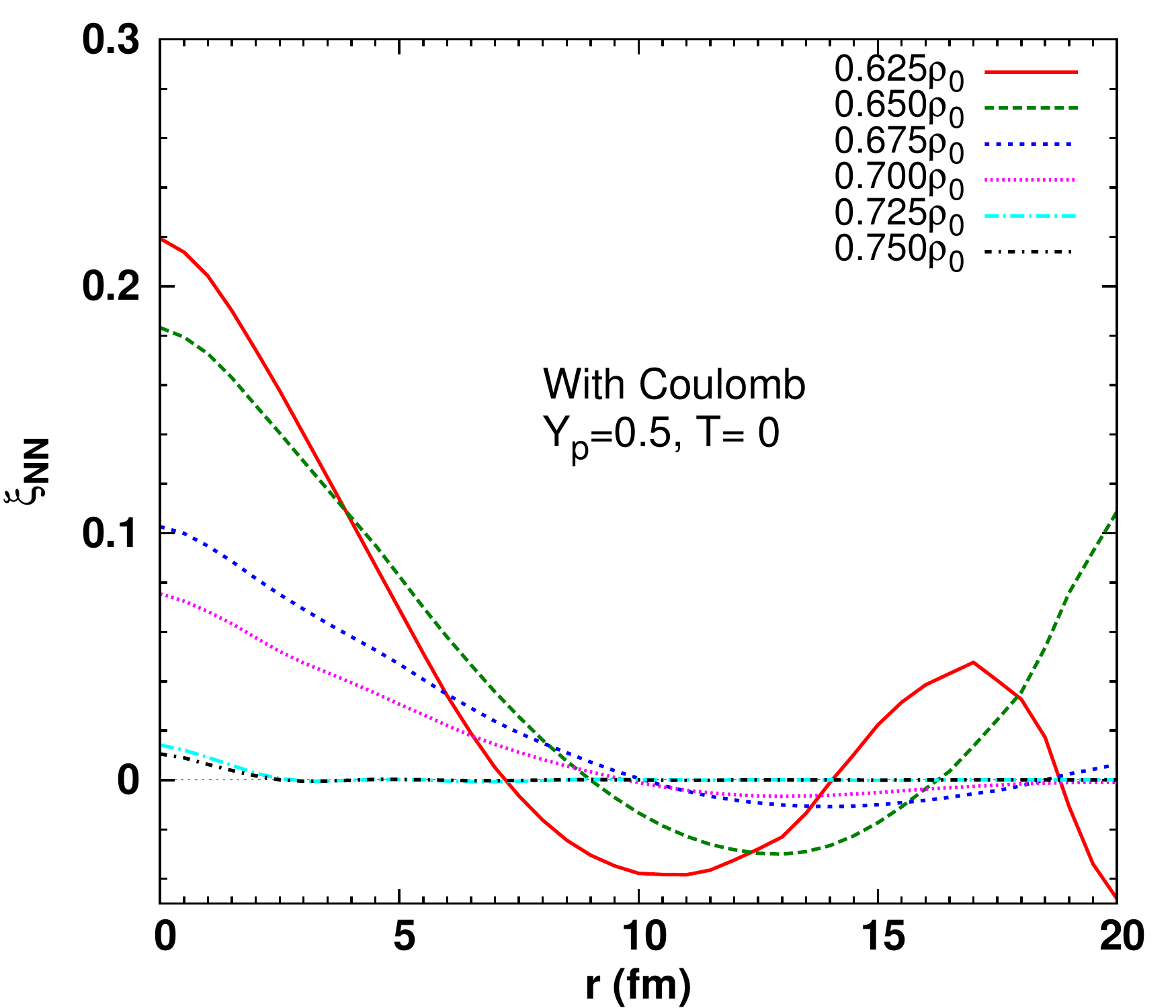}&
  \includegraphics[width=0.45\textwidth]{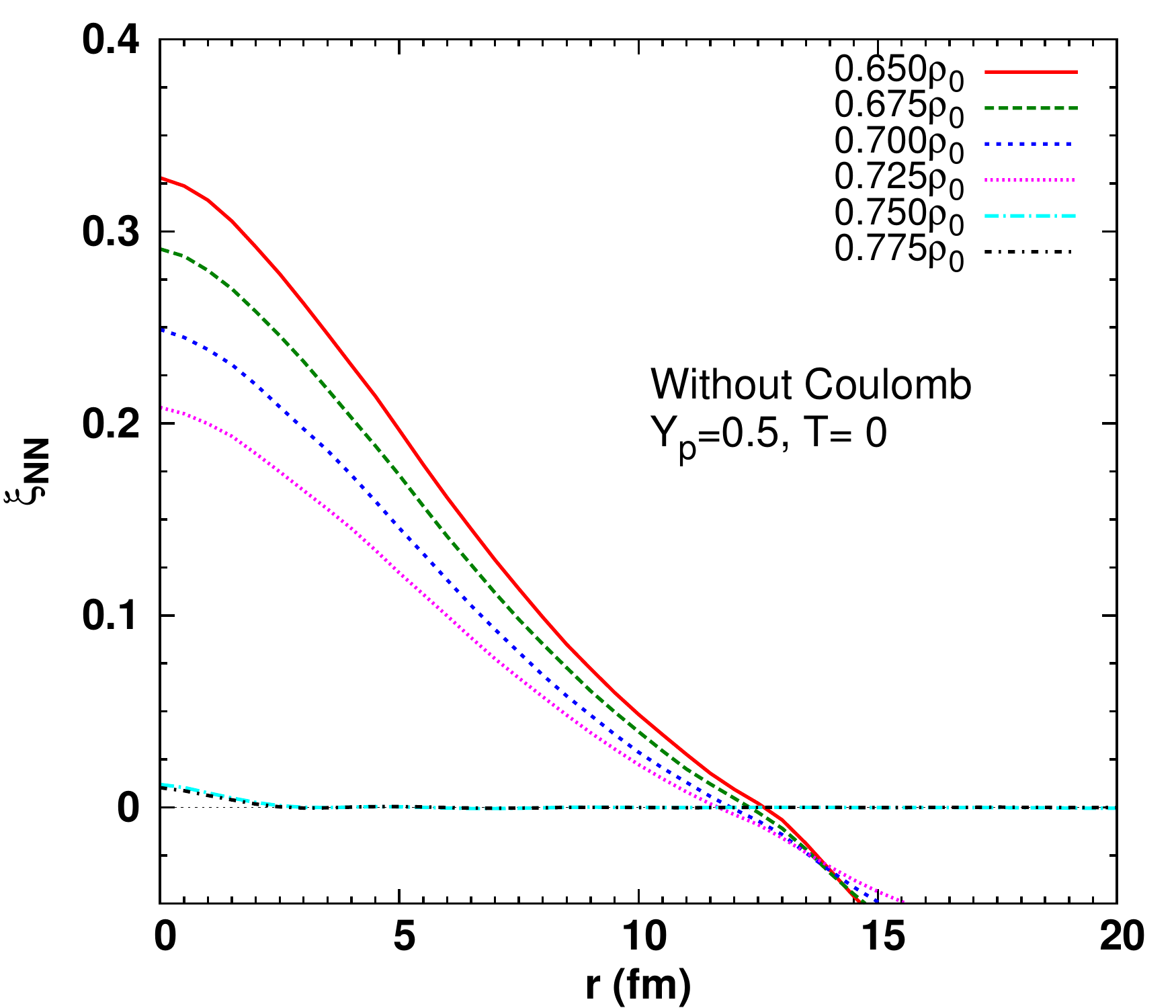}
 \end{tabular}
\caption{Two-point correlation function $\xi_{NN}$ around the liquid-gas phase transition region at $T=0$ with (left panel) and without (right panel)
Coulomb interaction for symmetric nuclear matter.}
\label{fig:correl_xp50_T0}
\end{figure}
In Fig. \ref{fig:correl_xp50_T0} we plot the two-point correlation functions for nucleons around the phase transition density
at $T=0$, with (left panel) and without (right panel) Coulomb interaction.  It is clear from the figure that the long-range correlation of the
nucleon distribution vanishes between 0.7-0.725$\rho_0$ when Coulomb interaction is considered. This value is in agreement with that of an earlier
calculation \cite{Watanabe03} with the same model. On the other hand, if the Coulomb interaction is not considered the long-range correlation 
disappears between 0.725-0.75$\rho_0$. In other words, the Coulomb interaction shifts the transition from the liquid phase to the gas phase to lower density.

\begin{figure}
 \begin{tabular}{cc}
  \includegraphics[width=0.45\textwidth]{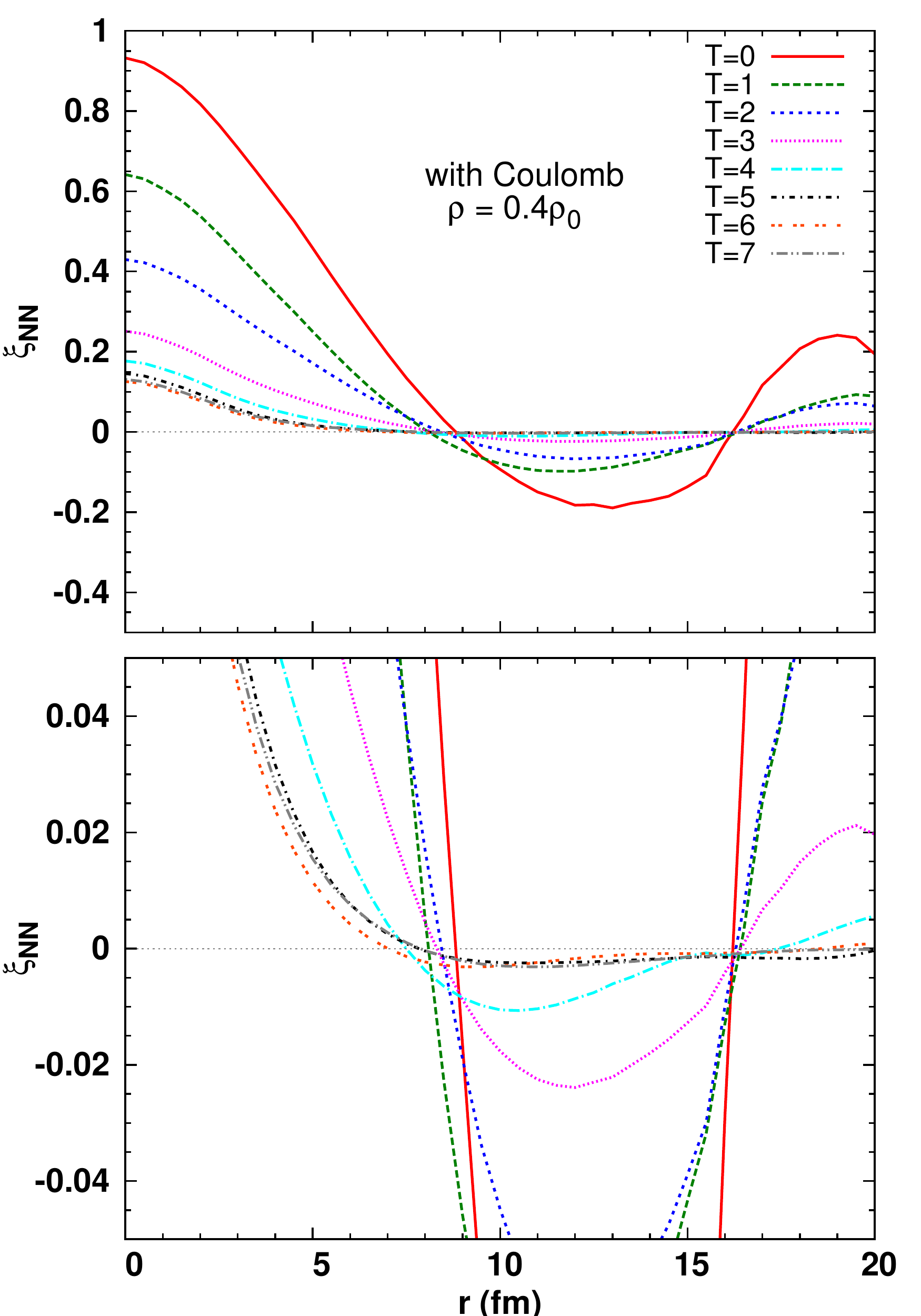}&
  \includegraphics[width=0.45\textwidth]{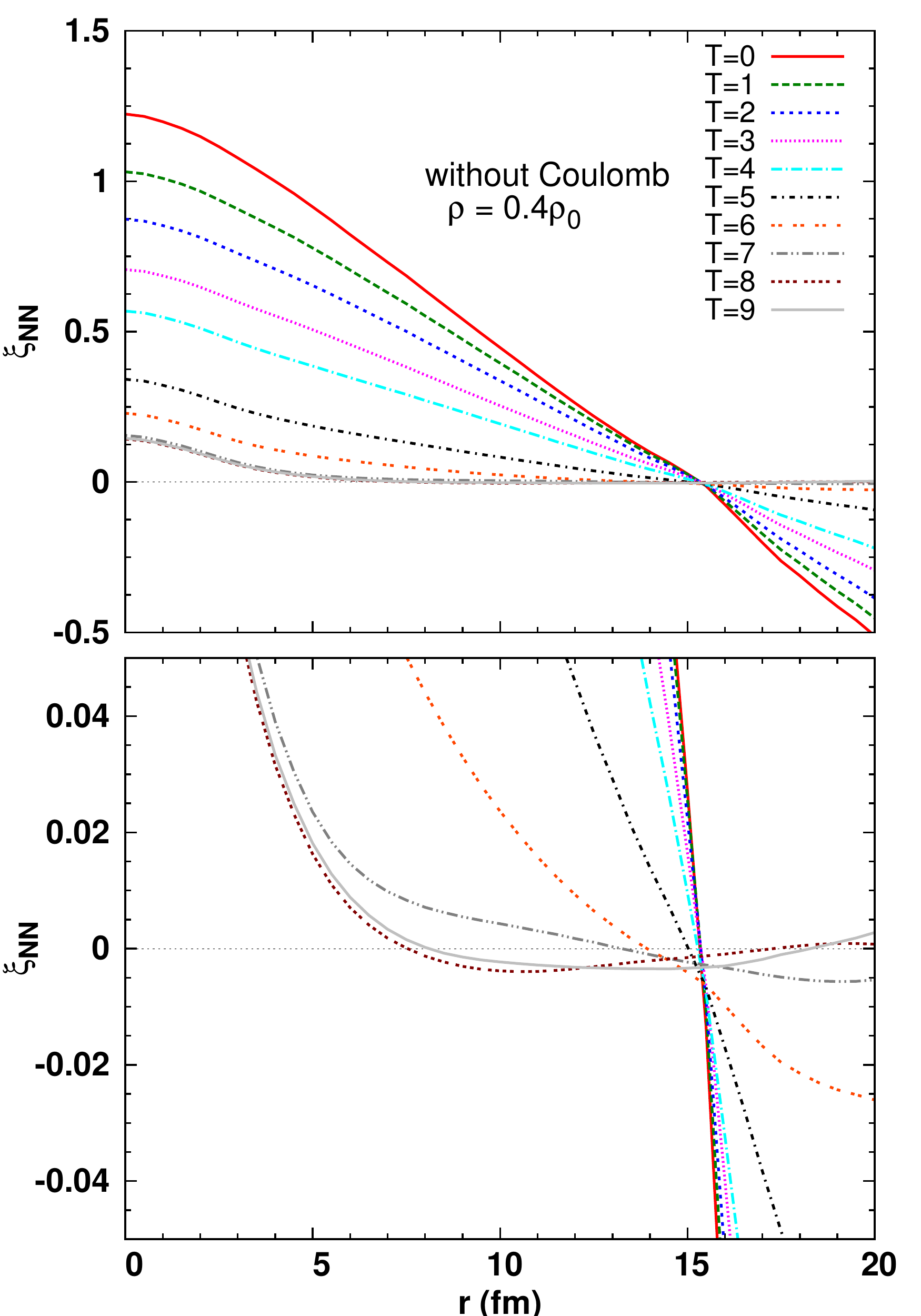}
 \end{tabular}
\caption{Two-point correlation function $\xi_{NN}$  at $\rho=0.4\rho_0$ with (left panel) and without (right panel) Coulomb interaction for symmetric
nuclear matter. The figures in the bottom panel are zoomed versions of the corresponding figures in the top panel}.
\label{fig:correl_p400nb_xp50}
\end{figure}

In Fig. \ref{fig:correl_p400nb_xp50} the two-point correlation function $\xi_{NN}$ is shown for the cases with (left panel)
and without (right panel) Coulomb interaction at $\rho=0.4\rho_0$, as a typical example. The figures in the bottom panel are the zoomed version of the
corresponding figures in the top panel. From the figures we find that although with increasing temperature the amplitude of $\xi_{NN}$ decreases, 
its first zero-point that corresponds to the size of the clusters doesn't change much. This behaviour was also seen in earlier
calculations \cite{Watanabe03, Watanabe04}. However, in absence of repulsive Coulomb
interaction between protons, nucleons are expected to form larger clusters. This is exactly seen here as
the first zero in $\xi_{NN}$ is reached at larger values of $r$ when the Coulomb interaction is not considered. A
Interestingly, one can also observe that the disappearance of long-range correlations that marks the transition from inhomogeneous to
homogeneous matter takes place between $T=4$ and 5 MeV in presence of the Coulomb interaction and between $T=6$ and 7 MeV without Coulomb
interaction. 

\begin{figure}
 \begin{tabular}{cc}
  \includegraphics[width=0.45\textwidth]{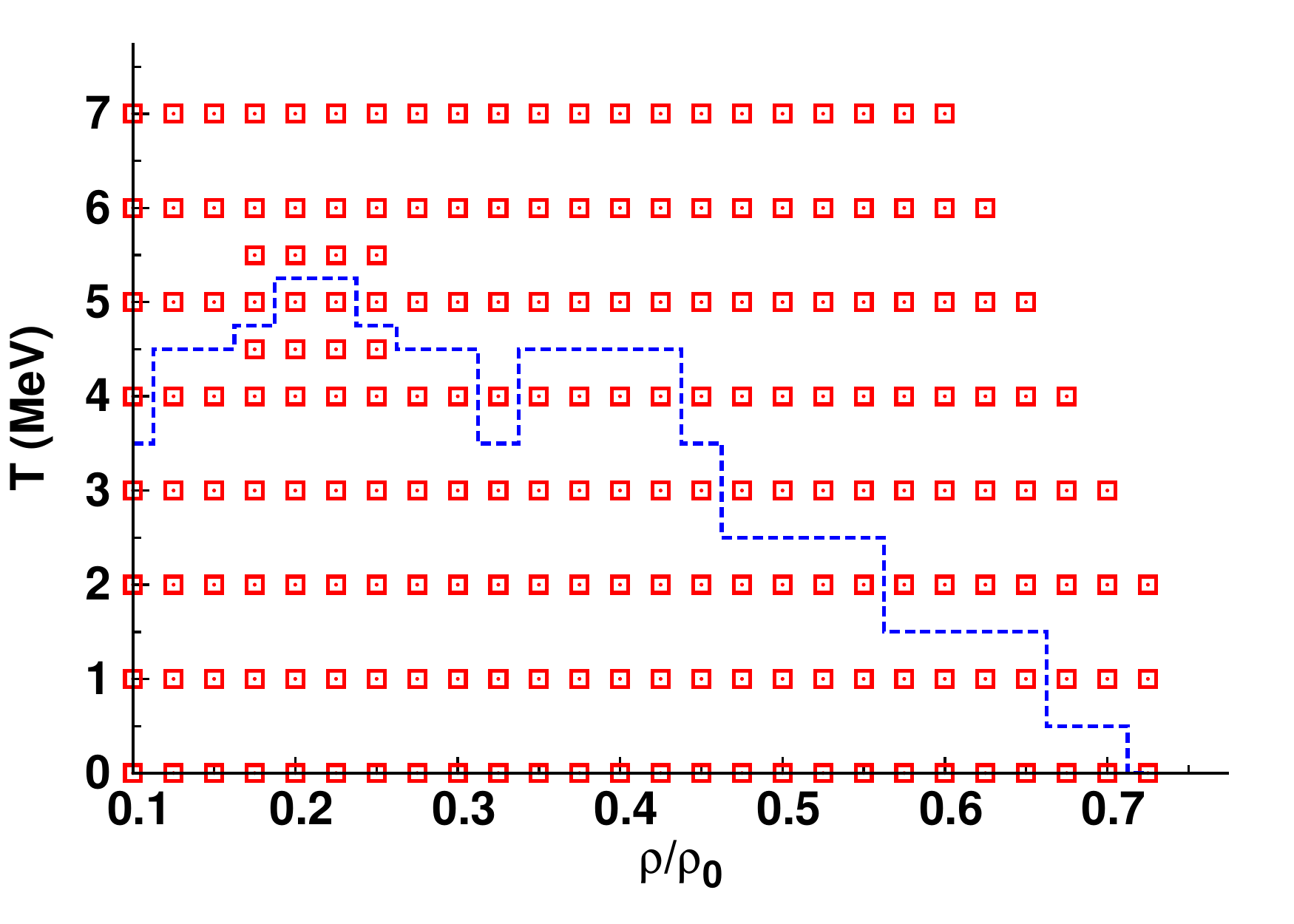}&
   \includegraphics[width=0.45\textwidth]{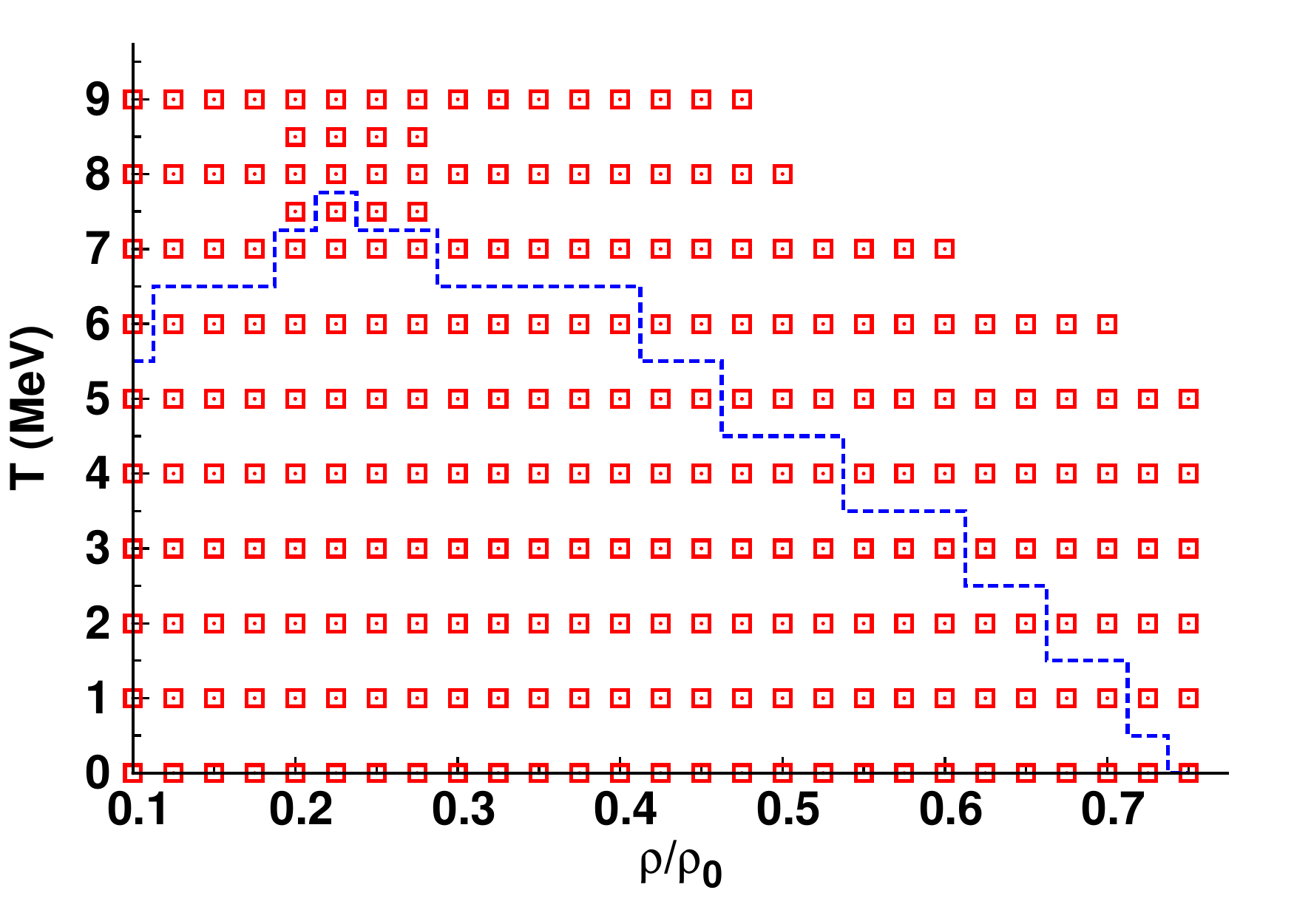}\\
 \end{tabular}
\caption{Phase diagram for symmetric nuclear matter with (left panel) and without (right panel) Coulomb interaction. 
The dashed line indicates the phase transition line.}
\label{fig:pd_xp50}. 
\end{figure}
Following the same procedure we determine the phase transition temperatures for the whole range of densities considered here for both with and
without Coulomb interaction. Accumulating all the results we obtain the phase diagrams shown in Fig. \ref{fig:pd_xp50}. Comparing the results
of two cases we can see that the phase transition temperature is always larger by $\gtrsim 1-2$ MeV for the case without Coulomb interaction
(right panel) than when including Coulomb  (left panel). From the phase diagrams it can also be observed that in the Coulomb case 
the critical end point of the liquid-gas phase transition is located at $\rho_c\simeq 0.225-0.25\rho_0$ and $T_c\gtrsim 5$ MeV. This value is similar to
the result obtained in an earlier calculation with the same QMD Hamiltonian \cite{Watanabe04}. On the other hand, the critical point is
located at $\rho_c\sim0.225$ and $T_c\gtrsim7.5$ MeV when Coulomb interaction is not taken into consideration. The important point to note is 
that the Coulomb interaction reduces the critical temperature $T_c$ by $\sim 2$ MeV but the critical density $\rho_c$ remains unchanged.
This behaviour was also seen in the calculation of Jaqaman {\it et al} \cite{Jaqaman84} for symmetric nuclear matter described by Skyrme interactions.

\subsection{Asymmetric nuclear matter with $Y_p=0.3$}
Next, we investigate the liquid-gas phase transition region for asymmetric nuclear matter with $Y_p=0.3$, a value typical for supernova environments.
We take 2048 nucleons (608 protons and 1440 neutrons) for this calculation. In this case also we calculate the two-point correlation functions 
to determine the phase boundary of the liquid-gas transition.
\begin{figure}
 \begin{tabular}{cc}
  \includegraphics[width=0.45\textwidth]{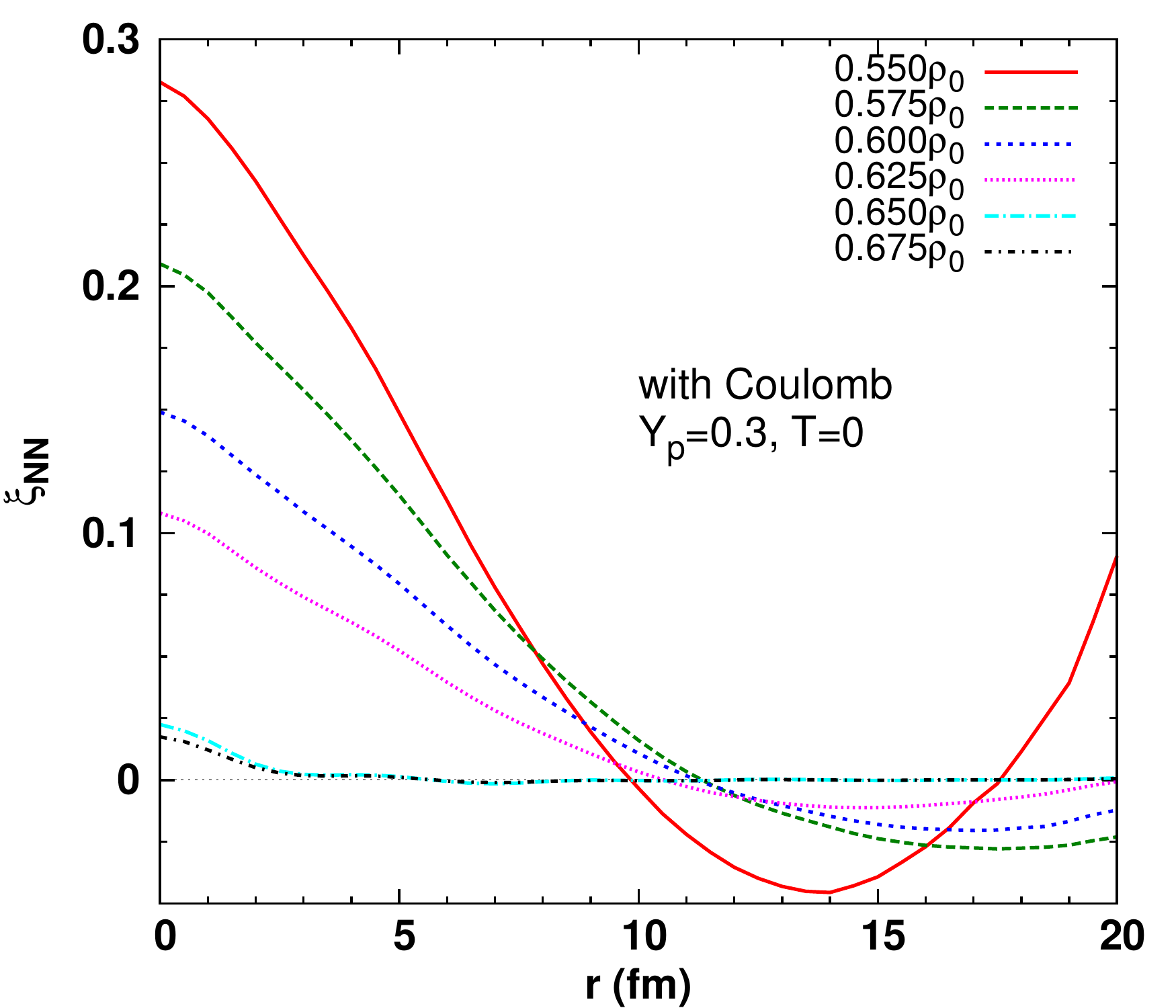}&
  \includegraphics[width=0.45\textwidth]{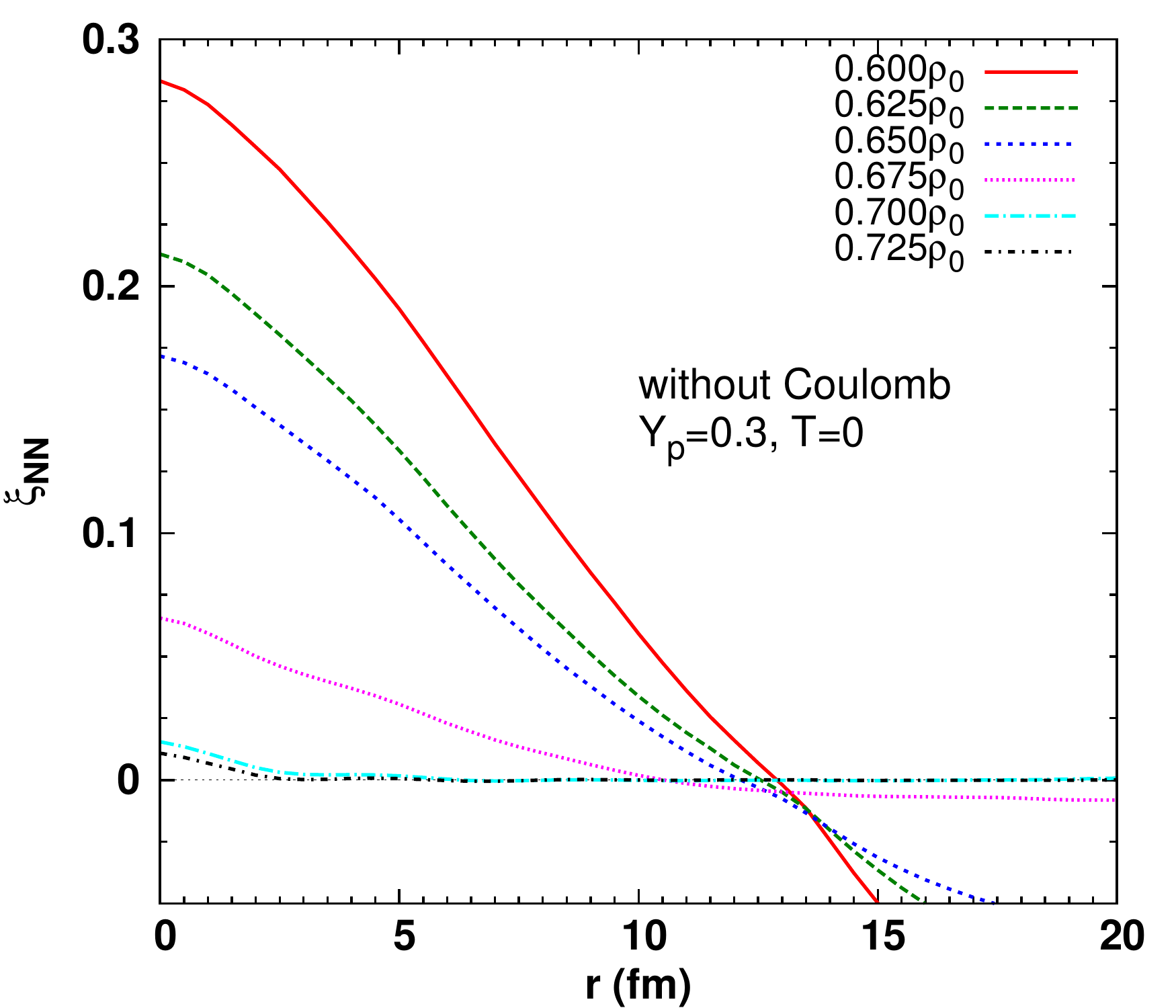}
 \end{tabular}
 \caption{Same as Fig. \ref{fig:correl_xp50_T0} but for asymmetric nuclear matter with $Y_p=0.3$}
\label{fig:correl_xp30_T0}
\end{figure}
In Fig. \ref{fig:correl_xp30_T0}, we plot the correlation function $\xi_{NN}$ at densities close to the phase transition region
for nuclear matter with $Y_p=0.3$ and $T=0$, with (left panel) and without (right panel) Coulomb interaction. As in the case of
symmetric matter here we also find that the Coulomb interaction decreases the liquid-gas transition density. With the Coulomb 
interaction the transition happens within the density range between $0.625$ and $0.65\rho_0$ (reported earlier in Ref. \cite{Nandi16}), whereas without Coulomb
interaction the same happens between $0.675$ and $0.7\rho_0$.

\begin{figure}
 \begin{tabular}{cc}
  \includegraphics[width=0.45\textwidth]{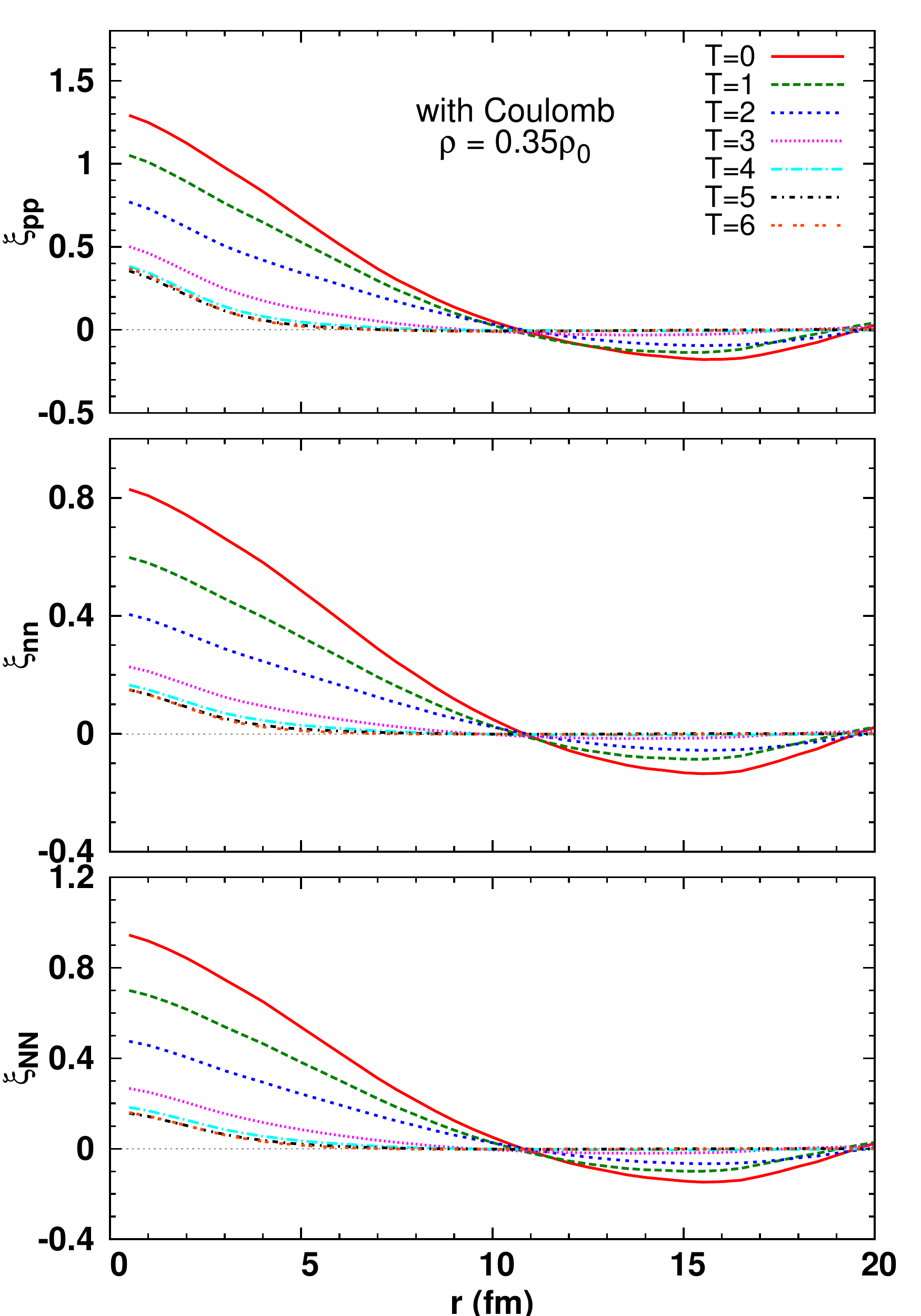}&
  \includegraphics[width=0.45\textwidth]{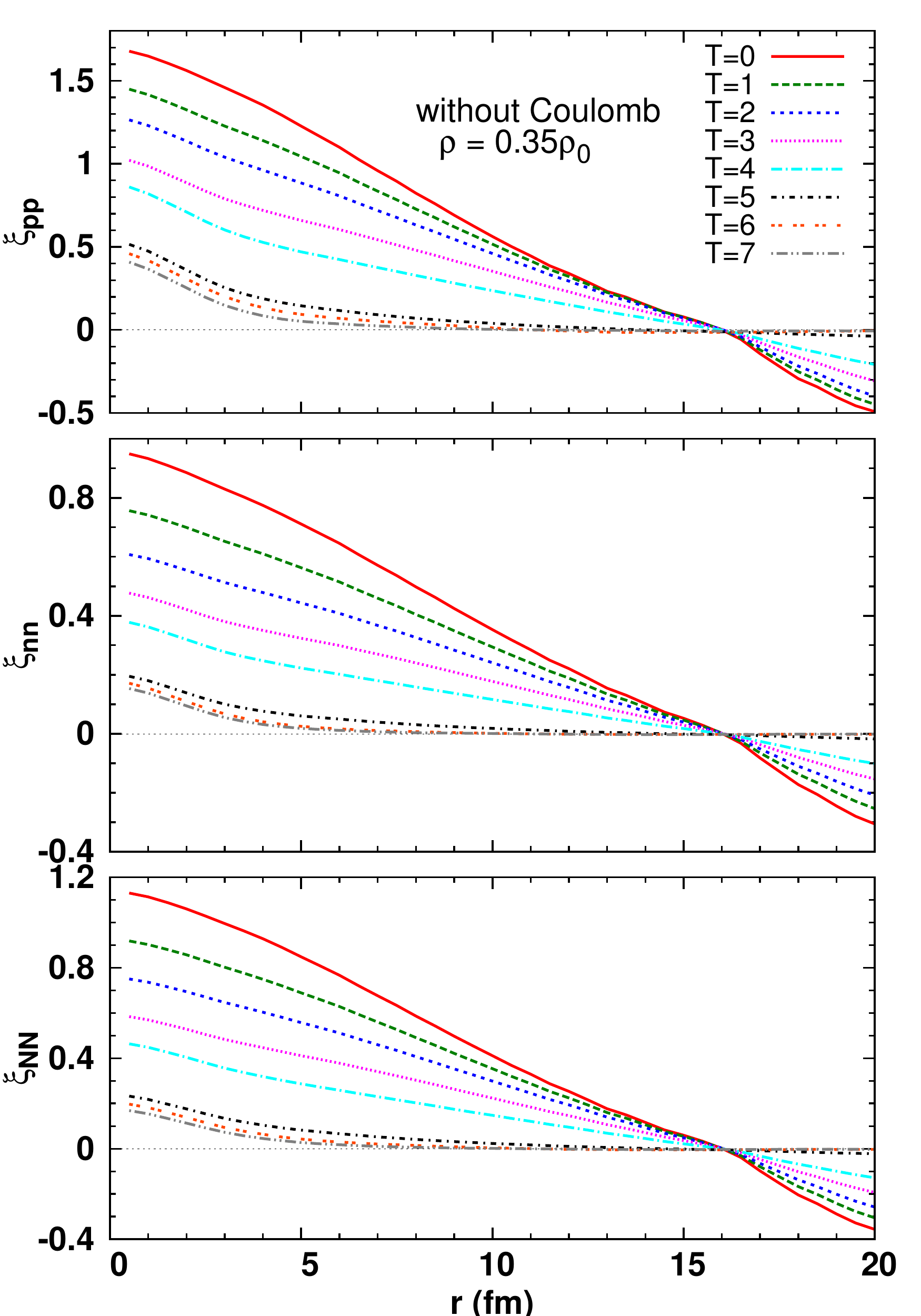}
 \end{tabular}
\caption{Two-point correlation functions at $\rho=0.35\rho_0$ with (left panel) and without (right panel) Coulomb interaction for asymmetric
nuclear matter with $Y_p=0.3$.}
\label{fig:correl_p350nb_xp30}
\end{figure}
We plot the two-point correlation functions $\xi_{pp},\xi_{nn}$ and $\xi_{NN}$  with and without considering the Coulomb interaction at a typical
example density $\rho=0.35\rho_0$, in Fig. \ref{fig:correl_p350nb_xp30}. The amplitudes of $\xi_{nn}$ are found to be lower than those of $\xi_{pp}$
due to the presence of uniformly distributed dripped neutrons. The higher amplitudes of $\xi_{ii}$ in absence of the Coulomb interaction point
to the fact that the particles are more clustered in this case as is also seen in Fig. \ref{fig:snapshots}.
Likewise in symmetric matter here also the first zero-point of all $\xi_{ii}$
does not change much with temperature. Moreover, the first zero-point of $\xi_{pp}$ coincide with the one of $\xi_{nn}$ for all temperatures
showing a strong correlation between the density fluctuations of neutrons and protons even at $Y_p=0.3$.
From the figure we similarly find that the phase transition line lies  between $T=3-4$ MeV and $T=5-6$ MeV for the cases with and without
Coulomb interaction, respectively.  

\begin{figure}
 \begin{tabular}{cc}
  \includegraphics[width=0.45\textwidth]{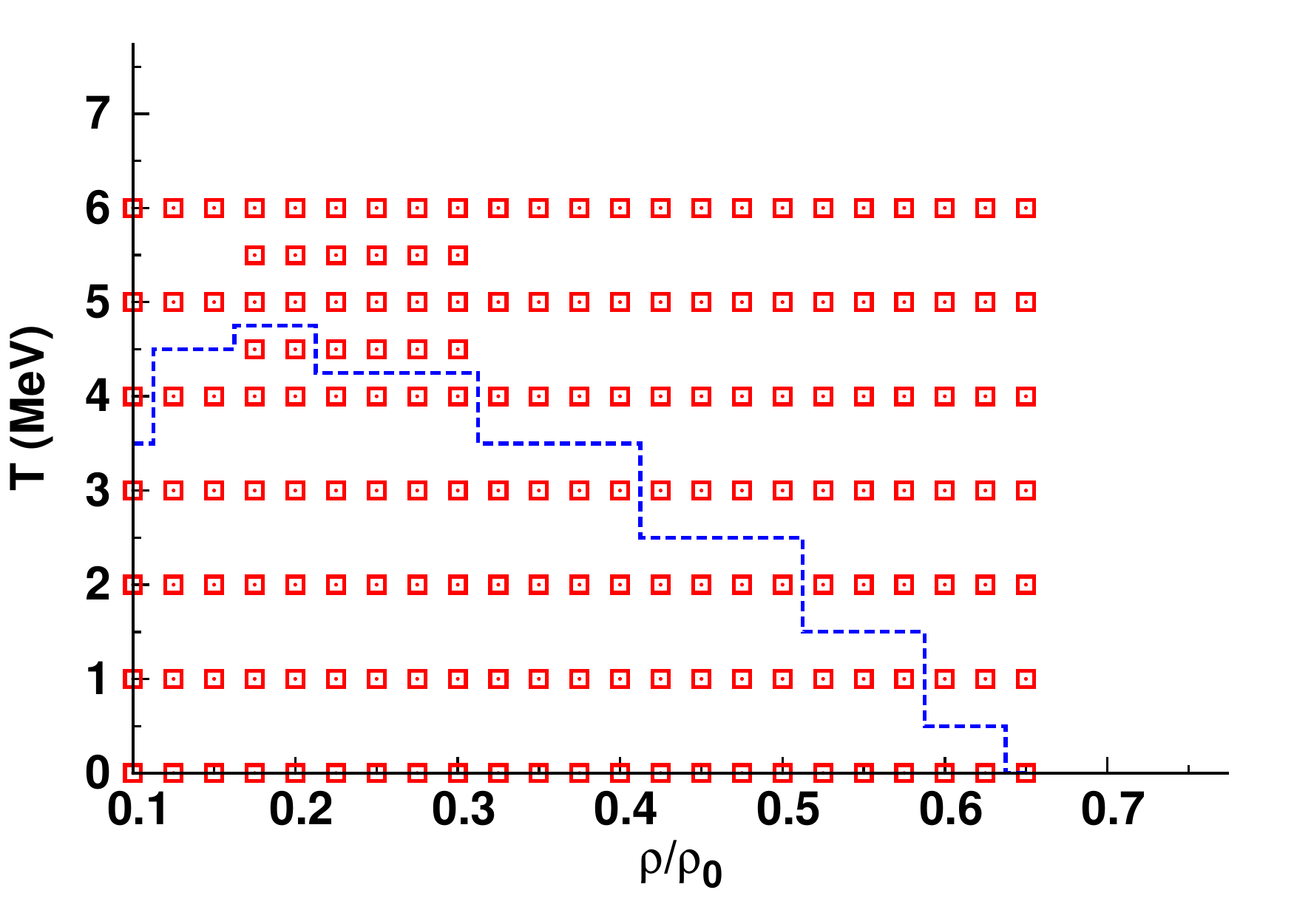}&
  \includegraphics[width=0.45\textwidth]{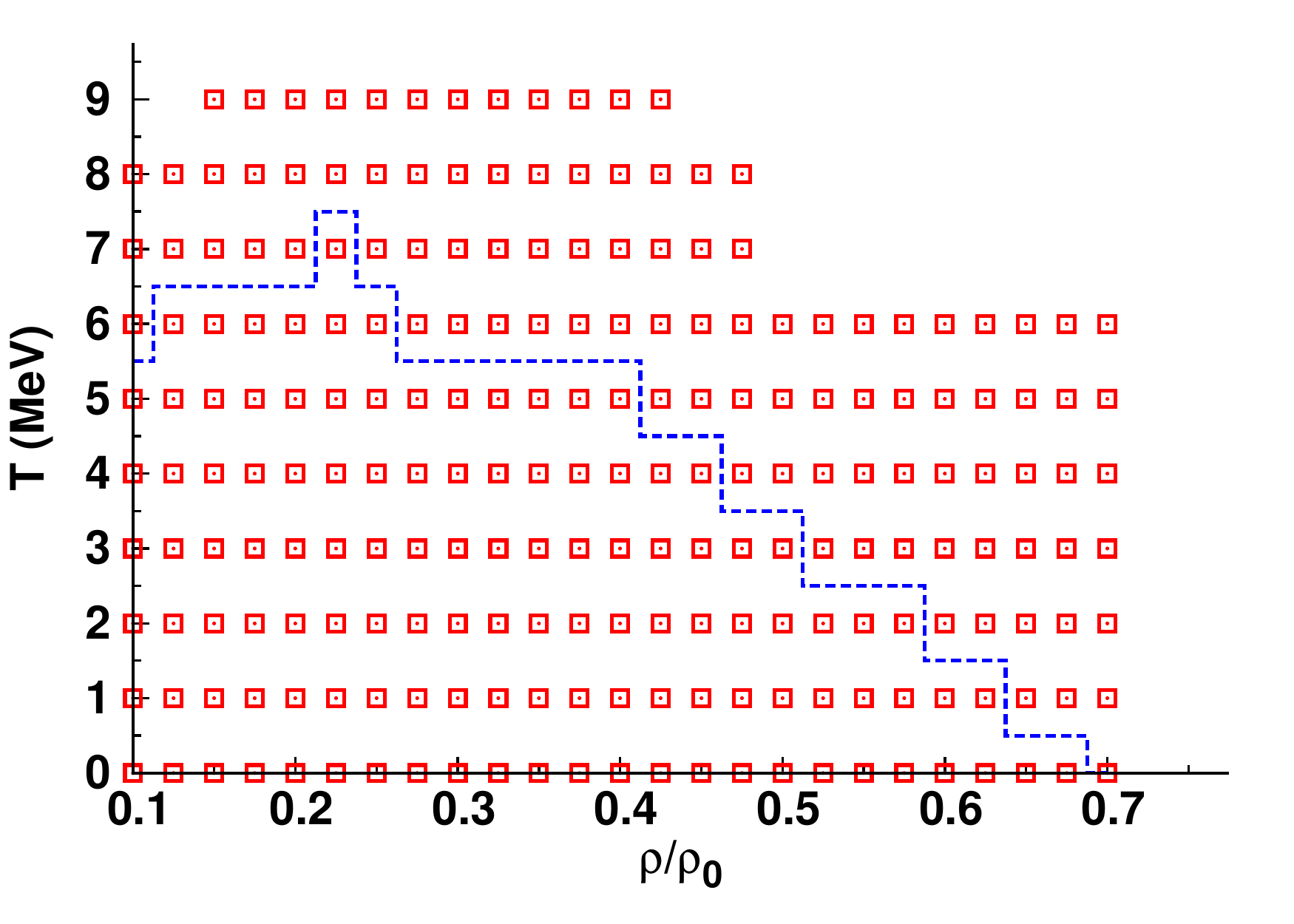}\\
 \end{tabular}
\caption{Same as Fig. \ref{fig:pd_xp50}, for $Y_p$=0.3}
\label{fig:pd_xp30}. 
\end{figure}
We continue the determination of the phase transition line for all other densities and obtain the phase diagram shown in Fig. \ref{fig:pd_xp30}.
When we do not consider the Coulomb interaction the transition temperature is always higher by $\gtrsim 1-2$ MeV compared to the case including
Coulomb interaction, analogously to the results for symmetric matter. The critical point of the transition is located at $T_c\gtrsim7$ MeV, $\rho_c \sim0.225\rho_0$ without
the Coulomb interaction and at $T_c\gtrsim4.5$, $\rho_c\sim0.2\rho_0$ with Coulomb. As for $Y_p = 0.5$ the critical 
density is not much affected by the Coulomb interaction but the critical temperature is decreased by $\gtrsim 3$ MeV.
A comparison of the results for $Y_p=0.5$ and 0.3 reveals that the critical point is similar in absence of Coulomb interactions.
However, including Coulomb the critical point gets shifted slightly to lower density and temperature.

\subsection{Asymmetric nuclear matter with $Y_p=0.1$}
Finally, we investigate the liquid-gas phase transition with and without Coulomb interaction for even more asymmetric
nuclear matter with $Y_p=0.1$ which is close to values relevant for the neutron star interior. In order to have enough number of protons
that play a crucial role in generating long-range correlations we use 16384 nucleons (1600 protons and 14784 neutrons) in our simulations for this case. The ground state
configurations are obtained following the same procedure as earlier. To determine the phase boundary of liquid-gas transition
we again calculate two-point correlation functions.
\begin{figure}
 \begin{tabular}{cc}
  \includegraphics[width=0.45\textwidth]{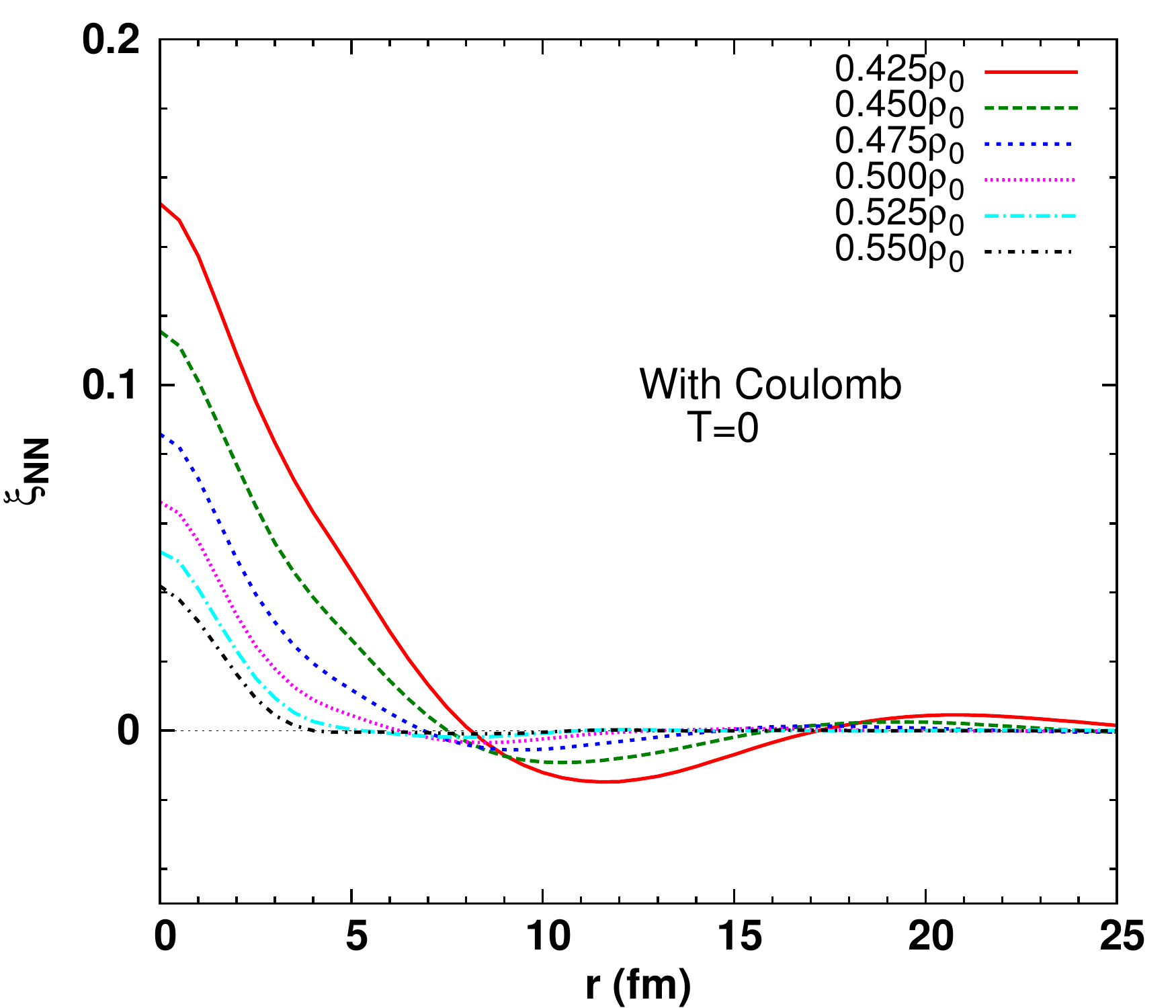}
  \includegraphics[width=0.45\textwidth]{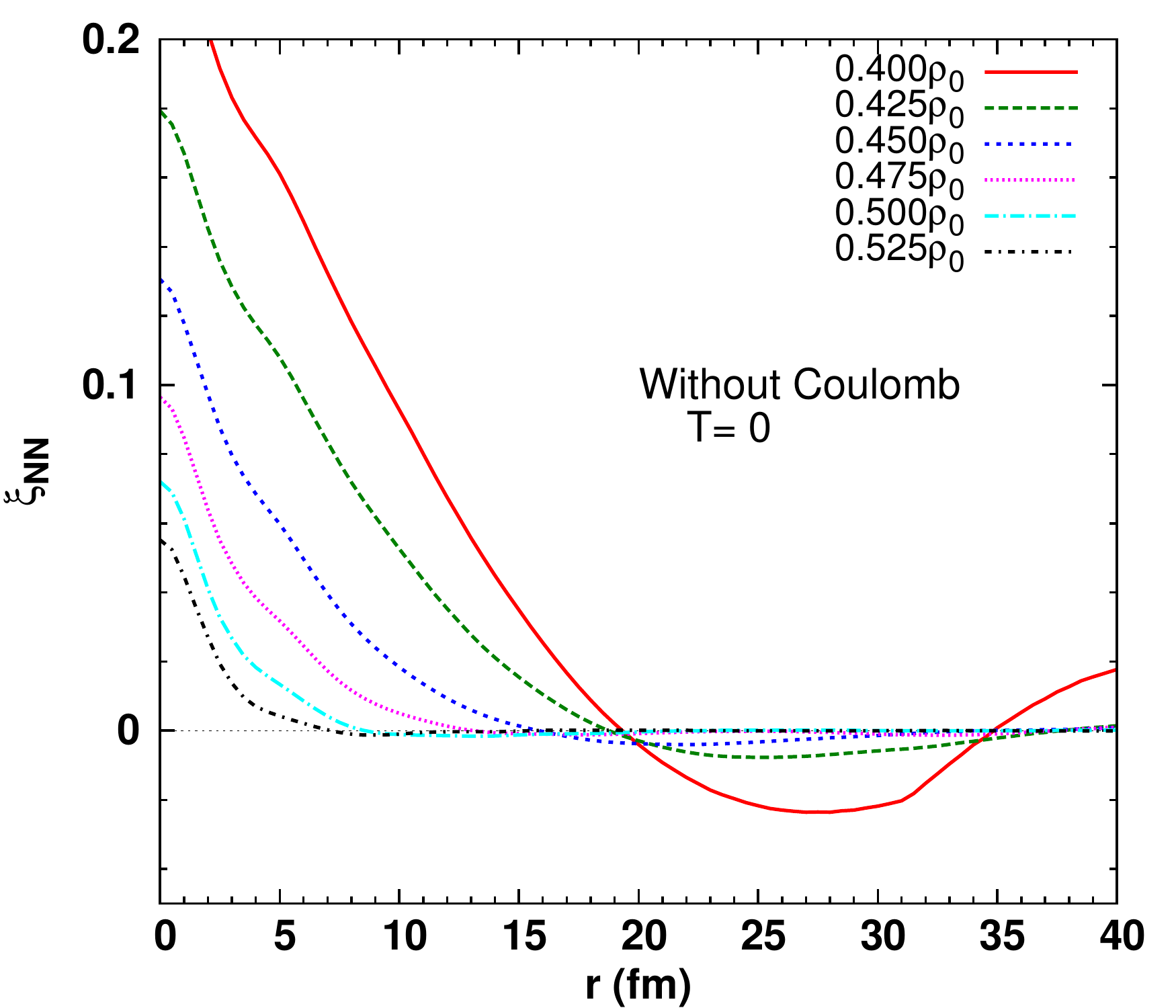}
 \end{tabular}
\caption{Two-point correlation as in Fig. \ref{fig:correl_xp30_T0} for $Y_p=0.1$.}
\label{fig:correl_xp10_T0}
\end{figure}
In Fig. \ref{fig:correl_xp10_T0}, we plot the correlation function $\xi_{NN}$ at densities close to the phase transition region
for nuclear matter with $Y_p=0.1$ and $T=0$, with (left panel) and without (right panel) Coulomb interaction.  Unlike the cases of 
$Y_p=0.5$ and $0.3$ here we find that the Coulomb interaction slightly increases the transition density. 
Including Coulomb  interactions the transition happens at a density range between $0.475$ and $0.5\rho_0$ whereas, without the Coulomb interaction 
this happens to be in the interval from $0.45$ to $0.475\rho_0$.

\begin{figure}
 \begin{tabular}{cc}
  \includegraphics[width=0.45\textwidth]{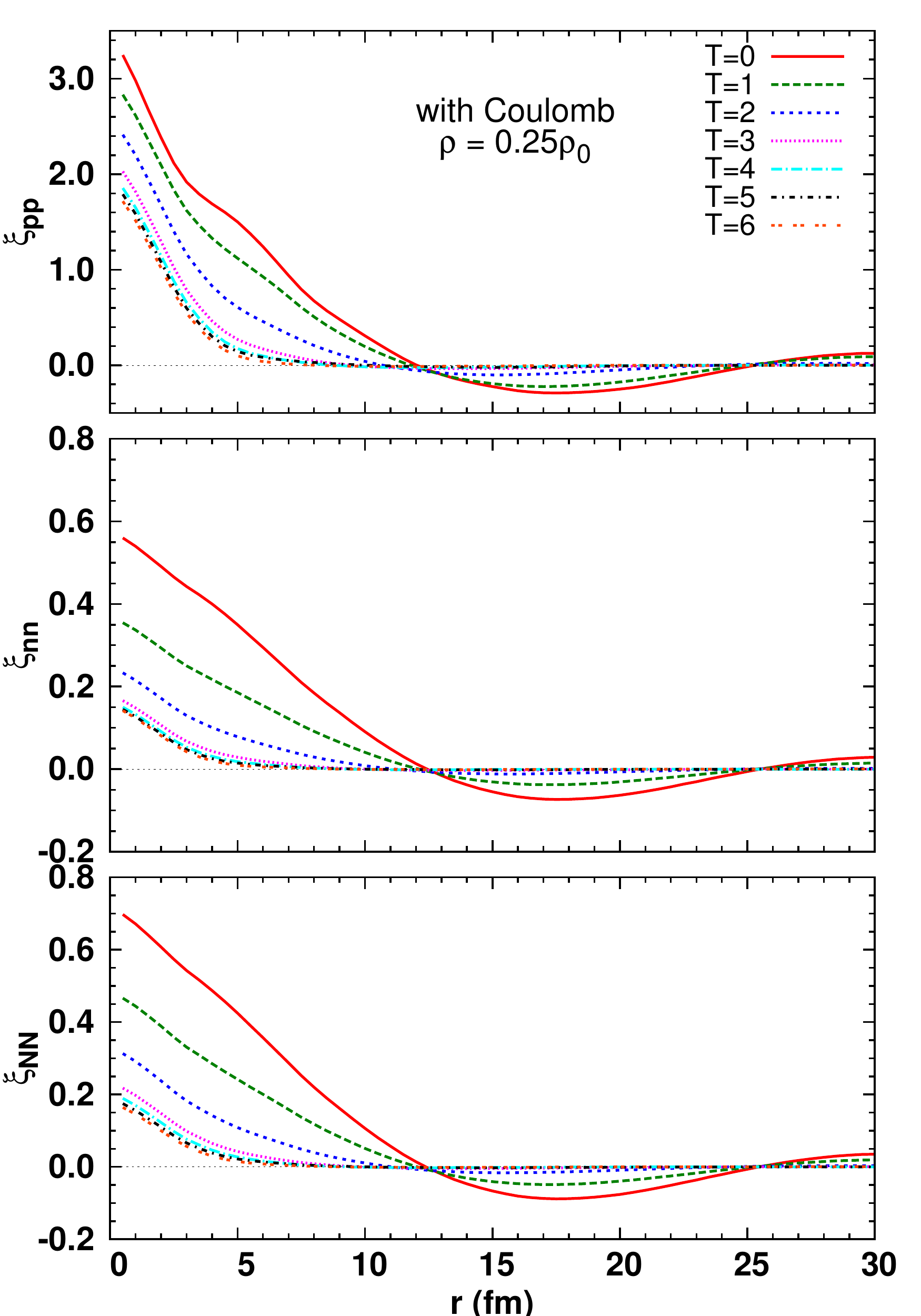}&
  \includegraphics[width=0.45\textwidth]{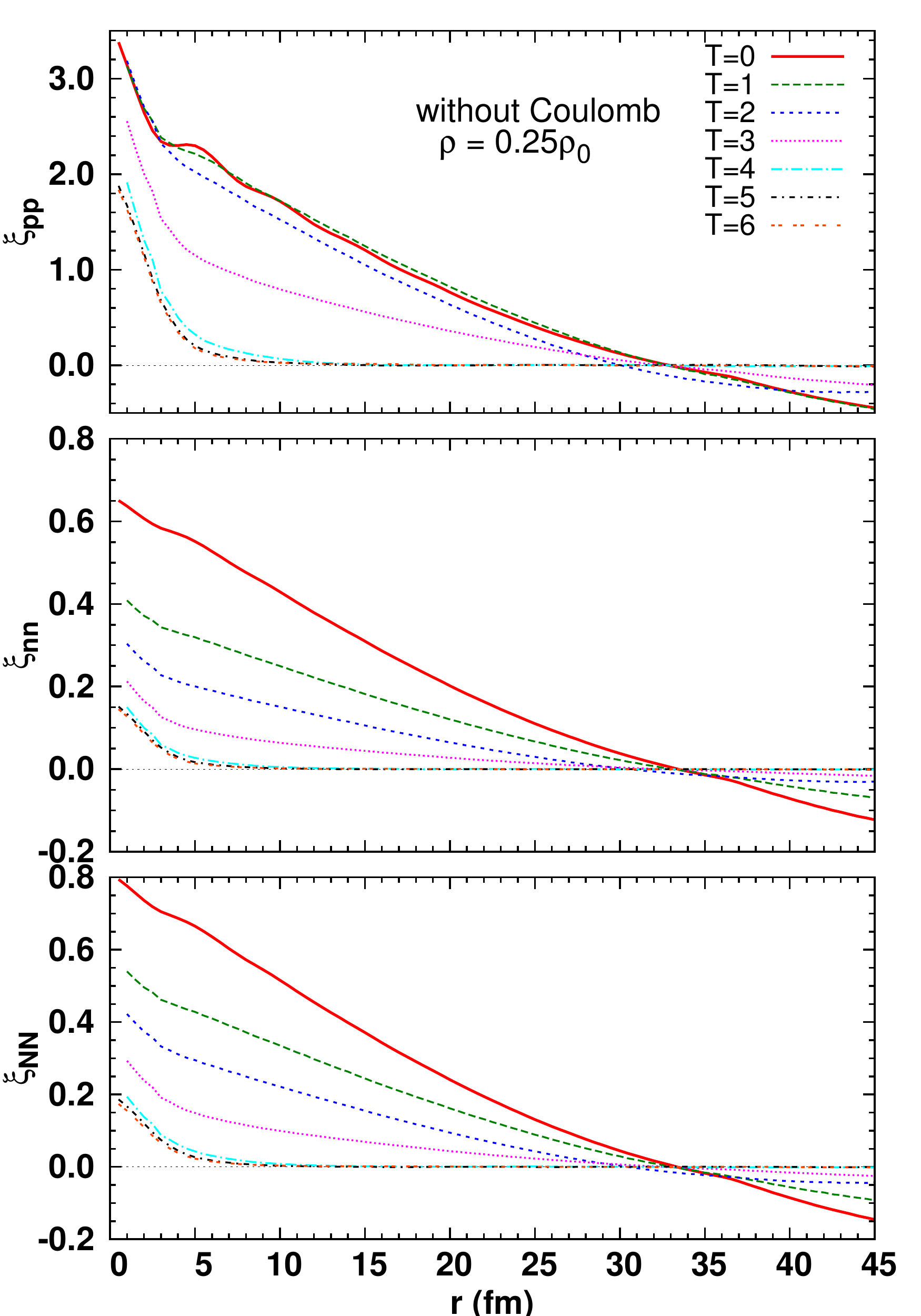}
 \end{tabular}
\caption{Two-point correlation functions at $\rho=0.25\rho_0$ with (left panel) and without (right panel) Coulomb interaction for asymmetric
nuclear matter for $Y_p=0.1$.}
\label{fig:correl_p250nb_xp10}
\end{figure}
Next, we plot different two-point correlation functions  with and without considering the Coulomb interaction at a typical
example density $\rho=0.25\rho_0$, in Fig. \ref{fig:correl_p250nb_xp10}. The difference between $\xi_{pp}$ and $\xi_{nn}$ is
even higher than that for $Y_p=0.3$, because the number of dripped neutrons also increases with decreasing $Y_p$. Even in this
highly asymmetric matter neutrons and protons are found to be highly correlated as the locations of first zero-point of $\xi_{ii}$
coincide. It is also seen from the figure that the long-range correlation vanishes between $T=2-3$ MeV and $T=3-4$ MeV for the
cases with and without the Coulomb interaction, respectively.

\begin{figure}
 \begin{tabular}{cc}
  \includegraphics[width=0.45\textwidth]{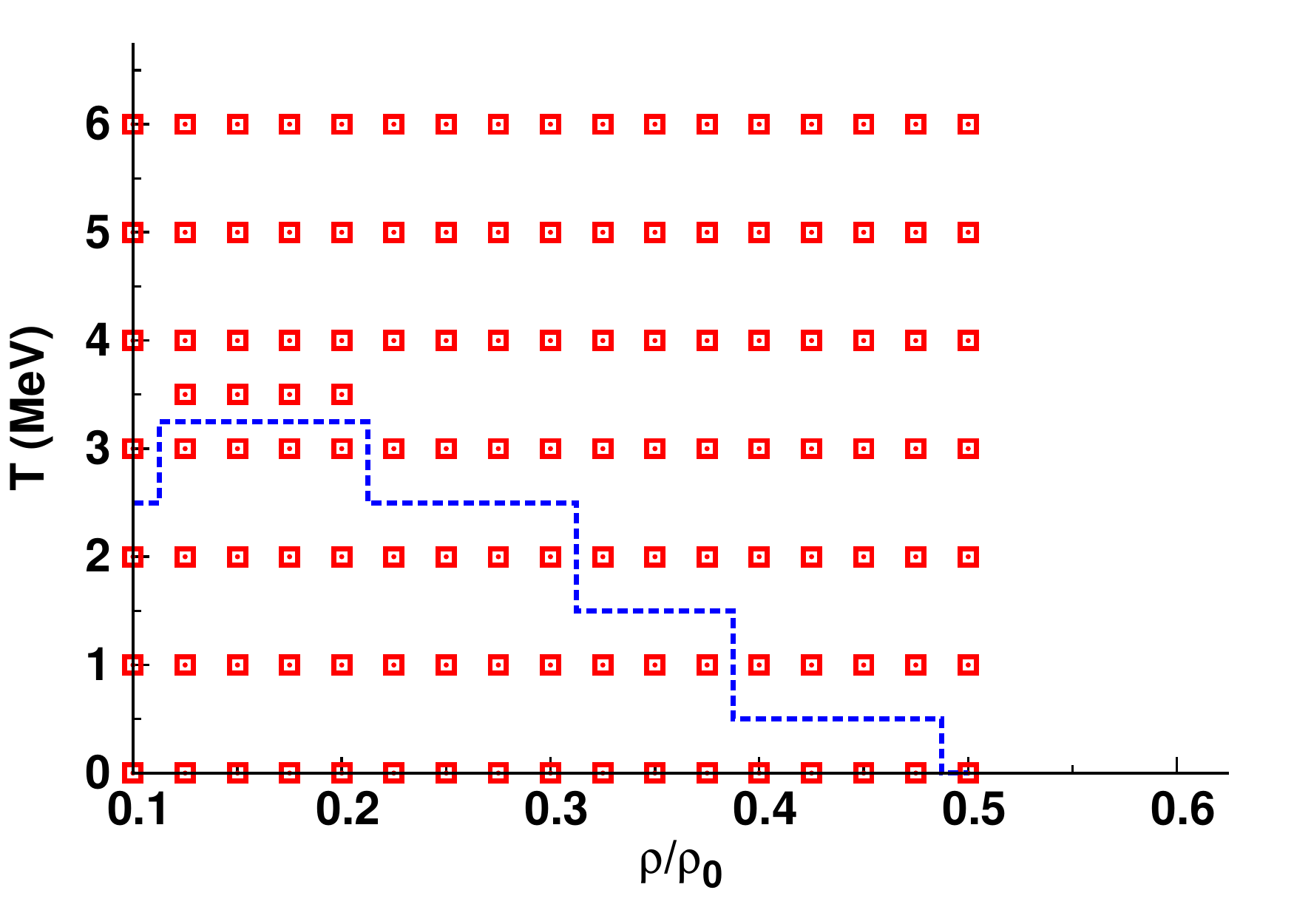}&
  \includegraphics[width=0.45\textwidth]{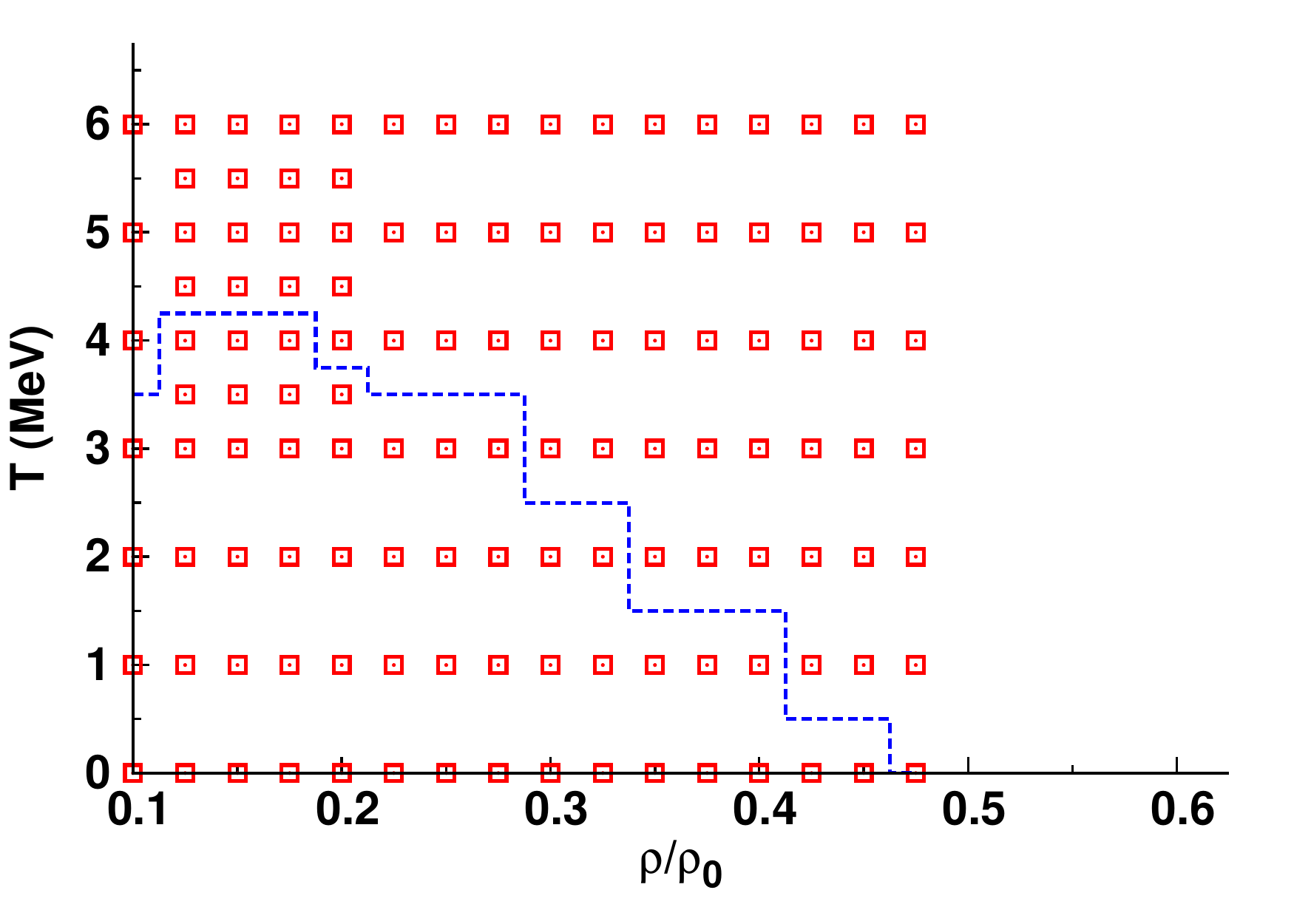}\\
 \end{tabular}
 \caption{Phase diagrams as in Fig. \ref{fig:correl_xp50_T0}, but for $Y_p=0.1$.}
\label{fig:pd_xp10}. 
\end{figure}
After determining the phase transition line for all other densities we obtain the phase diagram shown in Fig. \ref{fig:pd_xp10}.
From the figure we can see that the phase transition temperatures for the two cases differ by $T\sim 1$ MeV at low densities ($<0.3\rho_0$). 
However, at higher densities the difference vanishes.
The critical point of the transition is located at $T_c\gtrsim4$ MeV, $\rho_c \sim0.125\rho_0$ without
the Coulomb interaction and at $T_c\gtrsim3$, $\rho_c\sim0.125\rho_0$ with the Coulomb interaction. As in the cases of $Y_p=0.5$ and $0.3$
the critical density is not much affected by the Coulomb interaction but the critical temperature is decreased by only $\sim 1$ MeV in this case.
If the results of $Y_p=0.3$ and 0.1 are compared one can observe that the shift in critical point is larger when the Coulomb interaction is not
considered.

To investigate whether the reduction of the critical temperature in presence of the Coulomb interaction depends on the nuclear force, especially
on the surface energy we also perform simulations with another QMD model \cite{Chikazumi01} that includes a surface term . In this case we also
found that the Coulomb interaction reduces the critical temperature whereas the critical density remains largely unchanged. For this model, 
the critical point of the liquid-gas phase transition of nuclear matter with $Y_p=0.3$ is given by $T_c=10$ MeV, $\rho_c\sim 0.25\rho_0$, without
Coulomb and by $T_c=8$ MeV, $\rho_c\sim0.225\rho_0$, with Coulomb. The values of the critical temperature and density found here in presence of the 
Coulomb interaction are very similar to the values obtained by Sonoda {\it et al} \cite{Sonoda08} in an earlier study.
With increasing density and/or temperature the surface energy that depends
on the gradient of density across the surface, becomes smaller. Therefore, the reduction in critical temperature happens mainly due to the Coulomb 
energy and holds for all nuclear models.

\section{Summary and Conclusion}
We have investigated the effect of the Coulomb interaction on the liquid-gas phase transition of nuclear matter using molecular
dynamics simulations. We have performed simulations for a wide range of density and temperature with and without Coulomb interaction 
for this purpose. We have considered both symmetric nuclear matter, relevant for heavy-ion physics as well as asymmetric
matter with $Y_p=0.3$ and $0.1$, important for supernova and neutron star matter, respectively. To determine the phase transition
region we have calculated the two-point correlation functions of the fluctuations of nucleon densities. The temperatures at which the
transition from the liquid phase to the gas phase take place at various densities are obtained by determining the location where the long-range 
correlations vanish. We also determine the critical point of the liquid-gas phase transition of nuclear matter for all three $Y_p$s considered here.
We found that although the Coulomb interaction lowers the critical temperature by $\sim 2-3 $ MeV for nuclear matter with $Y_p=0.5$ and 0.3
and by $\sim 1$ MeV for $Y_p=0.1$, the critical density remains more or less unchanged. It could also be observed that the 
the densities at which the liquid-gas transition takes place at $T=0$, is higher if the Coulomb interaction is not considered
for the cases of $Y_p=0.5$ and 0.3. However, for $Y_p=0.1$, there is not much difference in the transition density.  For this
highly asymmetric matter the difference between the phase diagrams with and without Coulomb is much smaller than for the other two values 
of $Y_p$. This is the case because the Coulomb energy becomes less important for highly asymmetric matter.
We also showed that the main conclusion that the Coulomb interaction reduces the critical temperature but the critical density remain 
unchanged, is independent of nuclear model specifics.

Based on these findings we plan to investigate susceptibilities of particle numbers around the phase transition line and critical end-point, 
as such studies are directly related to the more general search for observable signals of structures in the phase diagram of strongly interacting
matter comparing to observables from heavy-ion collisions.

\end{document}